\documentclass[fleqn,usenatbib]{mnras}

\usepackage{newtxtext,newtxmath}

\usepackage[T1]{fontenc}

\DeclareRobustCommand{\VAN}[3]{#2}
\let\VANthebibliography\thebibliography
\def\thebibliography{\DeclareRobustCommand{\VAN}[3]{##3}\VANthebibliography}

\usepackage{graphicx}	
\usepackage{amsmath}	
\usepackage{multirow}

\usepackage{aas_macros}
\usepackage{multirow}
\usepackage{orcidlink}
\usepackage{physics}
\usepackage[version=4]{mhchem}

\usepackage{soul} 

\newcommand{\fugo}{f_{\ce{O2}}^{95}}
\newcommand{\fugs}{f_{\ce{S2}}^{95}}

\title[Atmospheres as a Window to Rocky Exoplanet Surfaces]{Atmospheres as a Window to Rocky Exoplanet Surfaces}

\author[X. Byrne \& O. Shorttle \& S. Jordan \& P. B. Rimmer]{
Xander Byrne
\orcidlink{0000-0001-9488-238X}$^{1}$\thanks{E-mail: ajnb3@ast.cam.ac.uk},
Oliver Shorttle
\orcidlink{0000-0002-8713-1446}$^{1,2}$,
Sean Jordan
\orcidlink{0000-0002-2828-0396}$^{1}$,
Paul B. Rimmer
\orcidlink{0000-0002-7180-081X}$^{2,3,4}$
\\
$^{1}$Institute of Astronomy, University of Cambridge, Madingley Rd, Cambridge CB3 0HA, United
Kingdom\\
$^{2}$Department of Earth Sciences, University of Cambridge, Downing St, Cambridge CB2 3EQ,
United Kingdom\\
$^{3}$Cavendish Laboratory, University of Cambridge, JJ Thomson Ave, Cambridge CB3 0HE, United
Kingdom\\
$^{4}$MRC Laboratory of Molecular Biology, Francis Crick Ave, Cambridge CB2 0QH, United Kingdom
}

\date{Accepted XXX. Received YYY; in original form ZZZ}

\pubyear{2023}

\begin{document}
\label{firstpage}
\pagerange{\pageref{firstpage}--\pageref{lastpage}}
\maketitle

\begin{abstract}
As the characterization of exoplanet atmospheres proceeds, providing insights into atmospheric chemistry and composition, a key question is how much deeper into the planet we might be able to see from its atmospheric properties alone. 
For small planets with modest atmospheres and equilibrium temperatures, the first layer below the atmosphere will be their rocky surface.  For such warm rocky planets, broadly Venus-like planets, the high temperatures and moderate pressures at the base of their atmospheres may enable thermochemical equilibrium between rock and gas.
This links the composition of the surface to that of the observable atmosphere.
Using an equilibrium chemistry code, we find a boundary in surface pressure--temperature space which simultaneously separates distinct mineralogical regimes and atmospheric regimes, potentially enabling inference of surface mineralogy from spectroscopic observations of the atmosphere.
Weak constraints on the surface pressure and temperature also emerge.
This regime boundary corresponds to conditions under which \ce{SO2} is oxidized and absorbed by calcium-bearing minerals in the crust, thus the two regimes reflect the sulphidation of the crust.
The existence of these atmospheric regimes for Venus-like planets is robust to plausible changes in the elemental composition.  Our results pave the way to the prospect of characterizing exoplanetary surfaces as new data for short period rocky planet atmospheres emerge.
\end{abstract}

\begin{keywords}
planets and satellites: atmospheres -- planets and satellites: composition -- planets and satellites: surfaces -- planets and satellites: terrestrial planets -- planets and satellites: individual: Venus
\end{keywords}



\section{Introduction
\label{intro}}

Since the discovery of the first exoplanet in orbit around a main sequence star \citep{mayor95}, over 5\,500 exoplanets have been discovered at the time of writing\footnote{
\url{https://exoplanets.nasa.gov/}
}.
High-resolution transmission spectroscopic facilities are enabling the increasingly routine characterization of exoplanetary atmospheres, with recent high-fidelity detections of \ce{CO2} \citep{jwstco2}, \ce{SO2} \citep{jwstso2}, and \ce{CH4} \citep{madhu23ch4} using \textit{JWST}.  As we gain greater insight into exoplanet atmospheres a key challenge is going to be seeing further into the planet, using the atmosphere as a window into the geological processes beneath.  

One key question to ask from atmospheric data is `what is the surface of the exoplanet?'.
There has been significant interest in identifying whether a planet has a discrete surface--atmosphere boundary at all \citep[e.g.,][]{hu2021_apjl,yu2021_apj,tsai2021_apjl}.
This is important where the bulk density and instellation of the planet leaves open multiple possibilities for the nature of the atmosphere-interior connection: whether it is a water ocean underlying a thick atmosphere \citep[e.g., Hycean worlds;][]{madhusudhan2021_apj}, a magma ocean \citep[e.g.,][]{lichtenberg2021_apj}, or a deep high pressure gas envelope (as with Jupiter and Saturn in our own solar system).
However, where the bulk density is constrained to be high, the equilibrium temperature is modest, and observations rule out thick atmospheres \citep[e.g., phase curve observations;][]{kreidberg2019_nature,zieba2023_nature}, then such planets may have atmospheres in contact with rock.
This opens up a regime of surface--atmosphere interaction of which we have direct analogues in the solar system's terrestrial planets.

\begin{figure*}
\includegraphics[width=\textwidth]{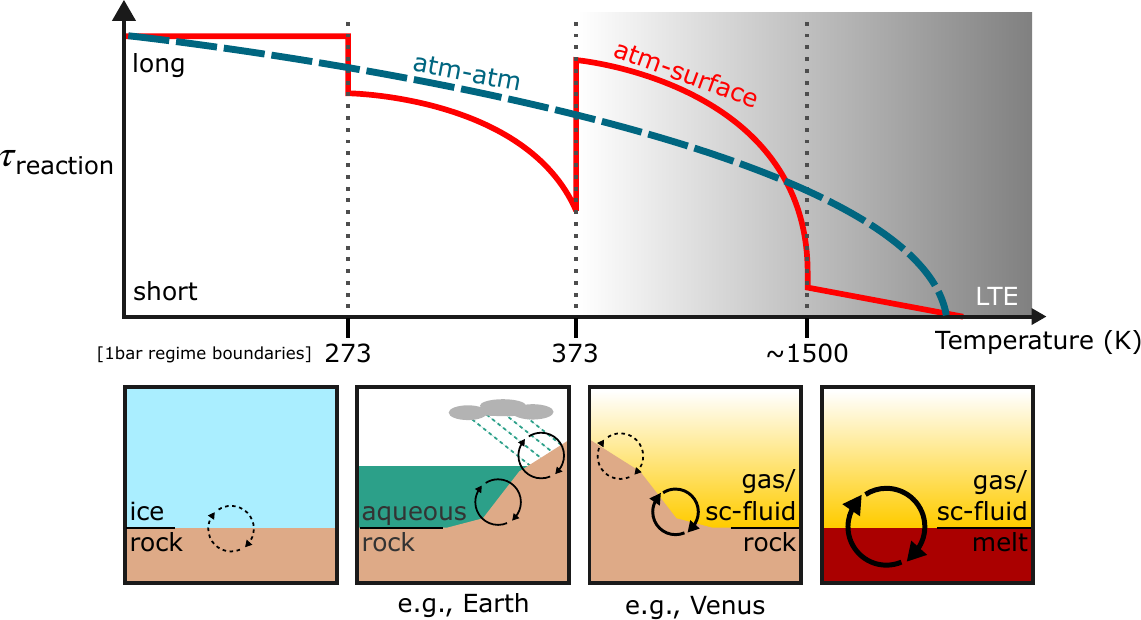}
\caption{
The timescales of surface--atmosphere interaction (red line) and atmospheric equilibration (blue dashed line) across different rocky planet climatic-geodynamic regimes.
A reference timescale is shown for pure gas-phase chemical equilibrium to be reached, which is a simple monotonic function of temperature (Arrhenius in nature).
Timescales of surface-atmosphere interaction on rocky planets deviate from this as a function of temperature as the climatic/geodynamic phases mediating the reactions change: schematics of these various planetary climatic-geodynamic regimes are shown at top, above their corresponding region of the reaction timescale graph. 
Cold planets feature slow solid-solid interactions between rock and ice layers (red line) and similarly cold atmospheres (dashed blue line).
Where the temperature allows liquid surface water, weathering processes efficiently couple the atmospheric, oceanic, and surface chemistry, albeit in a disequilibrium state.
Reaction rates in this regime increase with temperature as water temperature and increased rainfall drive more reaction \citep{walker1981_jgr,maher2014_science}.
In the absence of liquid water, the rate of interaction between the atmosphere and surface may sharply decrease.
However, at higher surface temperatures warm rocky planets such as Venus feature increasingly strong surface--atmosphere coupling.
In this regime thermochemical equilibrium is achievable potentially both in the lower atmosphere \citep{liggins23} and between the atmosphere and surface.
Very hot terrestrial worlds have molten surfaces, and there is likely efficient mass transfer (and therefore equilibration) between the surface and atmosphere.
}
\label{fig:regimes}
\end{figure*}

The mechanism of surface--atmosphere interaction depends primarily on the surface temperature, as depicted in Fig.~\ref{fig:regimes}.
On cool planets, such as Earth, liquid surface water facilitates geochemical interactions between a planet's lithospheric, oceanic, and atmospheric chemistry \citep[e.g.,][]{kite18}.
At intermediate temperatures -- like those at the surface of Venus ($\sim740\,$K), the so-called \textit{warm} exoplanet regime -- the surface and atmosphere may approach thermochemical equilibrium. In this latter case, the chemical species present at the base of the atmosphere and on the surface may then be prescribed by equilibrium thermodynamics, particularly at the high-temperature end of this regime.

Venus provides valuable insight into the potential nature of surface--atmosphere interaction on warm rocky worlds \citep[e.g., see][for a review]{fegley14}. 
After the first Venus missions found atmospheric temperatures in excess of $500\,\text{K}$, it was suggested that the planet's lithosphere is in equilibrium with its atmosphere \citep{mueller63}.
Subsequent lander missions measured surface temperatures of over $700\,\text{K}$, and pressures of over $90\,\text{bar}$ \citep{marov72}, consistent with conditions under which minerals metamorphose deep in the Earth's crust.
The findings of these missions fuelled the first theoretical considerations of surface--atmosphere interactions on Venus' surface \citep[e.g.,][]{lewis70, volkov86}, which indicated that a thermochemical equilibrium at the surface may regulate the abundances of major gases such as \ce{CO2} and \ce{SO2} in Venus' atmosphere.

Subsequent work has shown that Venus' atmospheric chemistry may mediate a large network of mineral equilibria, involving geochemically important minerals such as calcite (\ce{CaCO3}), anhydrite (\ce{CaSO4}), pyrite (\ce{FeS2}), and haematite (\ce{Fe2O3}) \citep[e.g.,][]{fegley97}.
The intractability of analytically analysing the vast number of important thermochemical equilibria on Venus and elsewhere requires equilibrium chemistry codes, e.g., \textsc{GGchem} \citep{woitke18} and \textsc{Perple\_X} \citep{connolly02}, to calculate the stable abundances of gas and mineral species at various temperatures and pressures.

\cite{lewis70} suggested that a surface--atmosphere equilibrium on Venus meant that the new knowledge of the planet's atmospheric chemistry could enable inferences on its surface mineralogy.
We posit that the same principle applies to exoplanets.
Exoplanetary atmospheres can be characterized in increasing detail with modern facilities, and for warm rocky exoplanets this offers as an exciting by-product the possibility of characterizing their surface mineralogy.
In turn, this surface mineralogy links to the composition and geodynamic history of planets, and is the key to understanding their wider geological history.
Such inferences could complement the use of emission spectroscopy in identifying surface mineralogy of exoplanets with a thin or absent atmosphere.

In this paper we investigate how the atmospheres of warm rocky exoplanets may provide information on their surface mineralogies. The paper is organized as follows.
In Section~\ref{methods}, we give an overview of relevant thermodynamics, and of the model we have used for our investigations.
In Section~\ref{grid}, we run the model over a broad 2D grid in surface pressure--temperature ($p_0$--$T_0$) space, identifying a boundary where the atmospheric chemistry and the surface mineralogy change abruptly and simultaneously.
Section~\ref{composition} explores the effects of altering the underlying elemental composition, to assess the generality of our findings.
In Section~\ref{discussion}, we discuss implications and caveats to our work, and Section~\ref{conclusion} provides a summary.

\section{Methods} \label{methods}

\begin{figure}
\includegraphics[width=0.46\textwidth]{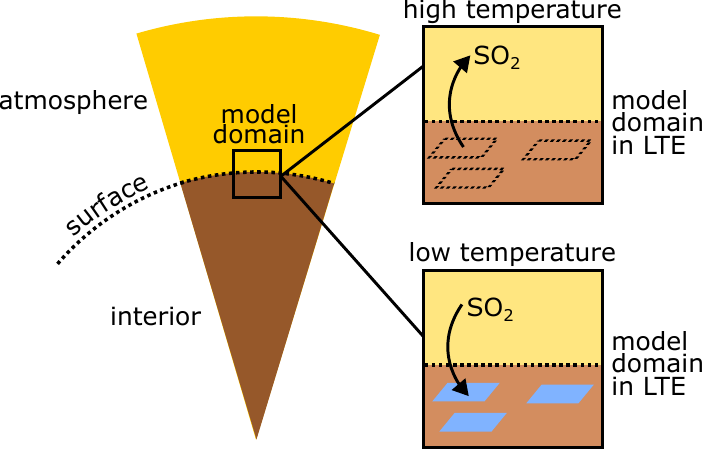}
\caption{
A schematic of the model setup.
We simulate conditions of lower atmosphere-surface interaction by calculating heterogeneous phase equilibria (gas-solid) in a single model domain.
At high temperatures components that can be present in solid form enter the gas phase, notably \ce{SO2} in our simulations.
At low temperature, those components can react back with the solid components into mineral form.
LTE = local thermochemical equilibrium.
}
\label{fig:model}
\end{figure}

The model setup is shown schematically in Fig. \ref{fig:model}.  We consider a system at fixed pressure and temperature that solves the heterogeneous solid--gas phase equilibria.
Depending on the temperature and pressure of the system, components (notably \ce{SO2}, as we will see in Section~\ref{grid}) will transition from the gas to solid phase.
Given the temperature range we consider in this study, major rock-forming oxides such as CaO, \ce{SiO2} etc. always remain in the solid phase.
Thereby, we can set the extent to which the solids (the `surface' of our model planet) buffer the chemistry of the gas (the `atmosphere') by changing the proportions of dominantly gas-phase to dominantly solid-phase elemental abundances.
A description of the approach to solving for thermodynamic equilibrium and the benchmarking of our calculations follows.

\begin{figure*}
\includegraphics[width=\textwidth]{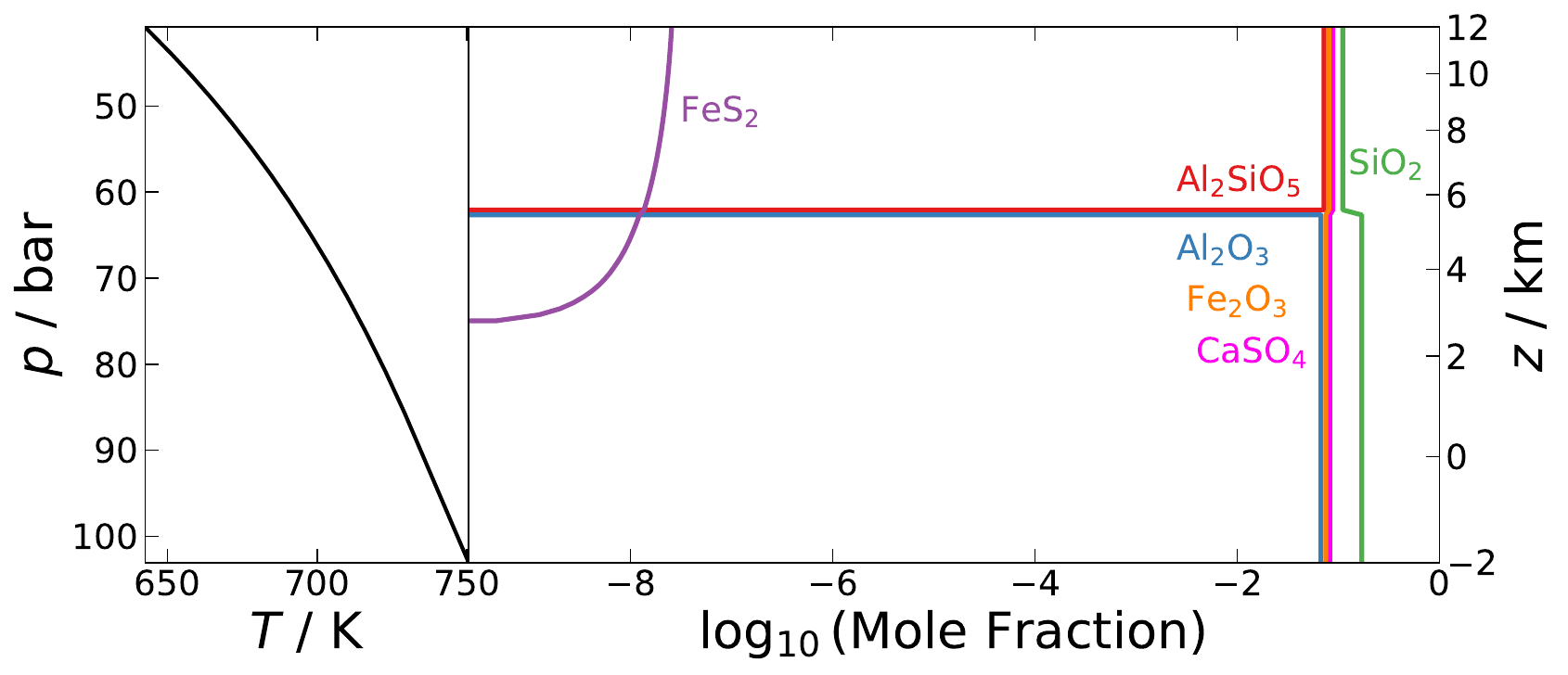}
\caption{
Venus' surface conditions and equilibrium abundances of selected minerals.
The left plot shows the $p$--$T$ profile on Venus, as provided by \textsc{GGchem};
the profile is not shown for altitudes higher than that of Skaði Mons (Venus' highest point) and is extrapolated down to the depth of the Diana Chasma (Venus' lowest point).
The right plot shows the equilibrium abundances of selected minerals, in terms of their mole fractions of the solid components.
The \ce{SiO2} line represents quartz itself, not the sum of all silicate minerals as is sometimes denoted by this.
The only two iron-bearing minerals present at any altitude are \ce{Fe2O3} and \ce{FeS2}, and the only two sulphur-bearing minerals are \ce{CaSO4} and \ce{FeS2}, consistent with previous thermochemical calculations \citep[e.g.,][]{rimmer21}.
The dark purple line shows that \ce{FeS2} only appears above an altitude of approximately $z=2.8\,\text{km}$, corresponding roughly to the altitude above which Venus shows high radar reflectivity \citep{pettengill82, pettengill96}.
We see that below an altitude of approximately $z=5.6\,\text{km}$, andalusite (\ce{Al2SiO5}; red line) decomposes into corundum (\ce{Al2O3}; blue) and quartz (green), the latter's abundance being slightly higher below this altitude.
}
\label{fig:venus}
\end{figure*}

\subsection{Thermodynamic modelling}
We first outline the requisite thermodynamics for calculating equilibrium mineral and gas abundances.
In thermochemical equilibrium, the set of volume mixing ratios (VMRs) $\{x_i\}$ of all gas species at a given pressure and temperature is that which minimizes the Gibbs free energy \citep[e.g.,][]{white58}:
\begin{equation}
\frac{G(p,T,\{x_i\})}{RT}
= \sum_i x_i \qty[
    \frac{G^\circ_{m,i}(T)}{RT}
    + \ln \qty(\frac{p}{p^\circ})
    + \ln x_i
],
\end{equation}
where $G^\circ_{m,i}(T)$ is the molar Gibbs free energy at the standard pressure $p^\circ=1\,\text{bar}$ and temperature $T$, and $p$ is the total pressure.
This minimization is subject to the constraints:
\begin{equation}
\sum_i x_i = 1;
\qquad
\sum_i a_{ij} x_i = b_j,
\end{equation}
where $a_{ij}$ is the number of atoms of element $j$ in species $i$ and $b_j$ is the total number density of atoms $j$.
The only free parameters are the pressure, temperature, and relative elemental abundances.
There are several codes which, given these parameters, calculate the equilibrium VMRs of a gas, but few also calculate the abundances of solid/liquid condensates that would be present in equilibrium with such a gas.
One code that does is \textsc{GGchem} \citep{woitke18}, which we use throughout this work.

It is worth justifying the use of this code to model surface--atmosphere equilibrium.
The initial use case of \textsc{GGchem} was to study the condensation sequence and the formation of astrophysical dust \citep{woitke18}.
To this end the code finds, under user-inputted conditions, the mixture of solids and gases that would be present at thermochemical equilibrium.
The solids are in this case envisioned as dust grains floating in the surrounding gas.
However, with the assumption of equilibrium, this is essentially equivalent to a surface in contact with an atmosphere.
Although the gas--solid interaction area is lower in the surface--atmosphere case, this is a kinetic barrier, not a thermodynamic one.
Provided the timescale for equilibration is significantly shorter than that of non-equilibrium chemical processes -- as is the case on Venus-like planets -- a surface mineralogy would be able to reach thermochemical equilibrium with an overlying atmosphere.
Indeed, it is expected that the surface mineralogy of Venus is in equilibrium with the lower atmosphere \citep[see, e.g.][]{mueller63, krasnopolsky81, volkov86, fegley97}
though this is to date unconfirmed by in situ measurements.
We show in Appendix~\ref{app:layerdepth}
that a layer of rock with a thickness of just $15\,\text{cm}$ suffices to react with the entire mass of the atmosphere; over this thickness crustal fractures allow effective contact with the atmosphere, facilitating equilibrium between the two.
Where the elemental composition of Venus is required, we use the procedure outlined in section 3.2 of \cite{rimmer21}, 
which uses \textit{Vega 2} measurements of oxide mass fractions on Venus' surface from \citet{surkov86}
to estimate elemental abundances.
The resulting composition is given in Table~\ref{tab:elements} in Appendix \ref{app:elements}.
We only consider the elements H, C, N, O, F, S, Cl, Fe, Mn, Si, Mg, Ca, Al, Na, K, and Ti, as well as He, Ne, and Ar.
As it is the surface minerals which provide the vast majority of the atoms in the surface--lower-atmosphere system, this surface composition effectively constitutes the overall composition of the equilibrium system.
Naturally, the underlying elemental abundance impacts the resulting chemical compositions of the surface and atmosphere; this is explored in Section~\ref{composition}.

\subsection{Benchmarking against modern Venus}
To verify the accuracy of \textsc{GGchem}'s calculations, we sought to validate the code against observations of Venus' altitude-dependent mineralogy.
The \textit{Pioneer Venus} and \textit{Magellan} missions found abnormally high radar reflectivity in the Venusian highlands \citep[$z\gtrsim2.6\,\text{km}$, relative to Venus' modal radius;][]{pettengill82, pettengill96}.
Venus' surface pressure and temperature both vary with altitude, resulting in the equilibrium abundances of surface minerals and gas species being altitude-dependent.
Hence, the highland radar reflectivity problem may be resolved by invoking some mineral that is thermodynamically stable only under Venus' high-altitude conditions, and not at low altitudes.
Various candidates for this mineral have been proposed, the most widely-accepted being pyrite \citep[e.g.,][]{klose92,kohler15,semprich20}, whose semiconductivity confers high radar reflectivity.
The presence of pyrite on the surface of the Venusian highlands is expected to be verified in the early 2030s by the \textit{VERITAS} \citep{veritas}, \textit{DAVINCI} \citep{davinci} and \textit{EnVision} \citep{envision} missions.

\textsc{GGchem} was run on Venus' surface conditions and elemental composition over the altitude range $-2\,\text{km}$ to $+12\,\text{km}$, spanning Venus' surface hypsometry \citep{fegley14}.
The full surface composition at $z=0\,\text{km}$ is given in Table~\ref{tab:majormins}.
We corroborate the results of \citet{rimmer21}, finding a solid composition of mainly enstatite (\ce{MgSiO3}), anorthite (\ce{CaAl2Si2O8}), and albite (\ce{NaAlSi3O8}).
The only iron- or sulphur-bearing minerals at $z=0\,\text{km}$ are \ce{Fe2O3} and \ce{CaSO4}.
The $p$--$T$ profile and the calculated abundances of selected minerals are shown in Figure~\ref{fig:venus}, including all iron- and sulphur-bearing minerals.
We see that pyrite becomes stable only when the pressure falls below $75\,\text{bar}$, corresponding to altitudes above roughly $2.8\,\text{km}$.
At lower altitudes, all of the iron is present as \ce{Fe2O3}; only a small fraction of the iron is transferred from \ce{Fe2O3} to \ce{FeS2}.
The only other significant change in mineralogy with altitude is the conversion of corundum (\ce{Al2O3}) and some quartz (\ce{SiO2}) to andalusite (\ce{Al2SiO5}), however this occurs at much higher altitudes ($z\approx 5.6\,\text{km}$) and andalusite is not a semiconductor, so its appearance is unrelated to the radar reflectivity observations.
The agreement between \textsc{GGchem}'s calculations and the likely source of Venus' highland radar reflectivity is evidence of the accuracy of the thermodynamic model under these conditions, justifying its application to Venus-like exoplanets.

\begin{table*}
    \centering
    \begin{tabular}{ccccccccccc}
    \hline
    \multicolumn{11}{c}{Solid Composition (mass fraction)}\\
    \hline
    \ce{MgSiO3} & \ce{CaAl2Si2O8} & \ce{NaAlSi3O8} & \ce{Fe2O3} & \ce{CaSO4} & \ce{SiO2} & \ce{Al2O3} & \ce{KAlSi3O8} & \ce{Mn3Al2Si3O12} & \ce{TiO2} & \ce{MgF2} \\
    29.7\% & 21.7\% & 17.6\% & 8.9\% & 8.3\% & 7.6\% & 5.1\% & 0.6\% & 0.3\% & 0.2\% & trace\\
    \hline
    \end{tabular}
    \caption{
        Equilibrium surface composition at $z=0\,\text{km}$ on Venus.
        Our results are identical to table 3 of \citet{rimmer21} at the precision shown.
    }
    \label{tab:majormins}
\end{table*}

\section{Coupled surface and atmosphere chemical regimes}
\label{grid}

\subsection{Gas and mineral chemistry}

\begin{figure*}
\centering
\includegraphics[width=\textwidth]{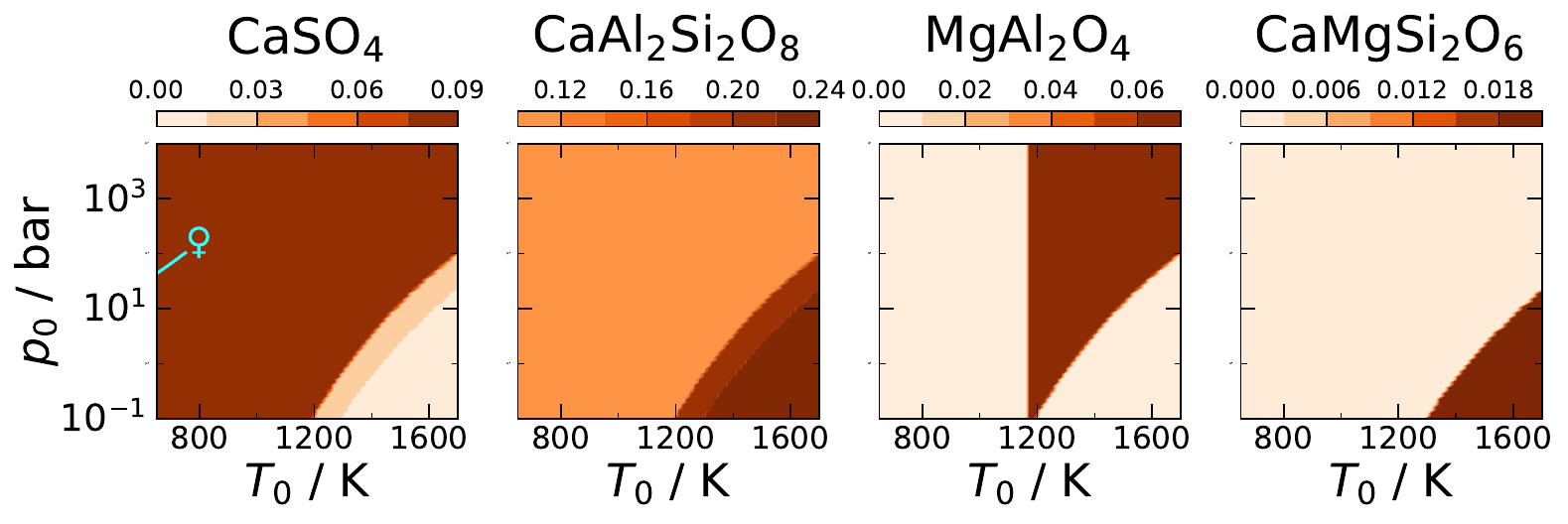}
\includegraphics[width=\textwidth]{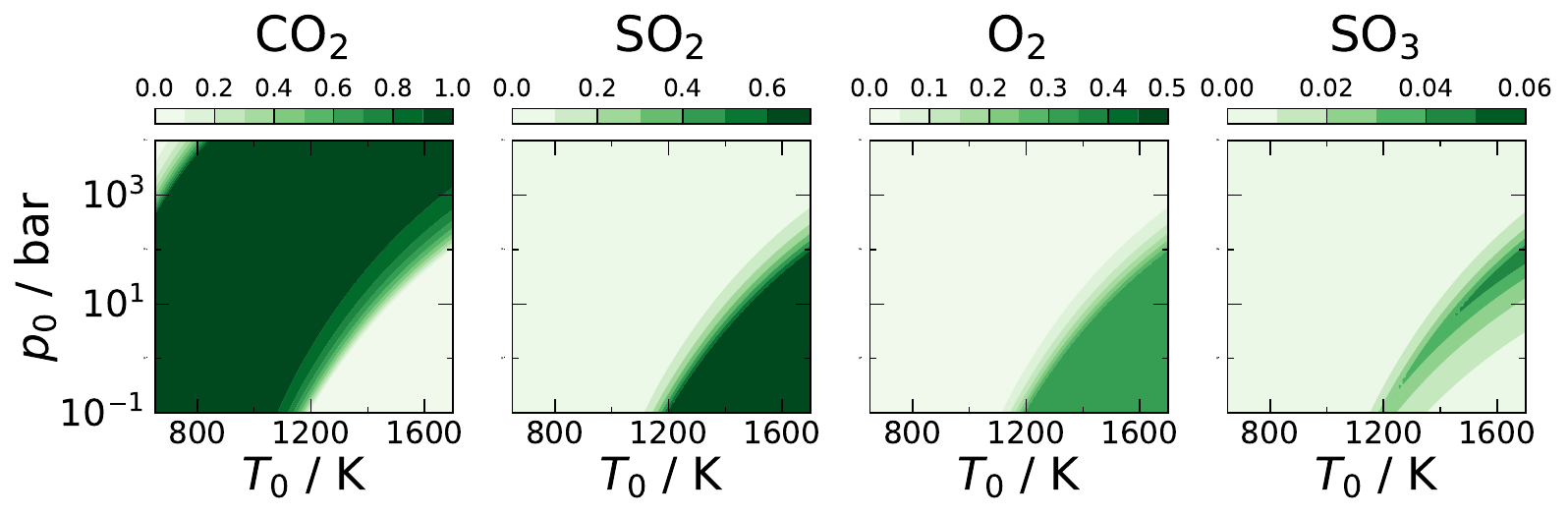}
\caption{
Equilibrium abundances of selected minerals and gases on Venus-composition exoplanets.
Equilibrium abundances of four major minerals are shown, in terms of mole fraction of the condensates, in the top row.
The surface conditions on Venus are shown by the short blue track in the first panel.
There is a clear boundary in the abundance patterns of these minerals, across which their abundances change abruptly.
The second row, shows the abundances of four major gases presented as VMRs, and demonstrates that the mineralogical boundary also demarcates regimes in the atmospheric chemistry.
The high-$T_0$, low-$p_0$ corner is dominated by a mixture of \ce{SO2} and \ce{O2}, with \ce{SO3} a minor component.
The rest of the atmospheres are dominated by \ce{CO2}.
}
\label{fig:grids}
\end{figure*}

The altitudinal variations in pressure and temperature on Venus' surface correspond to a 1D track in $p_0$--$T_0$ space.
To explore a wider range of conditions that may exist on warm rocky exoplanets, \textsc{GGchem} was run over a grid in $p_0$--$T_0$ space, with $p_0$ and $T_0$ distributed logarithmically in the respective ranges $[10^{-1}, 10^{4}]\,\text{bar}$ and $[650,1700]\,\text{K}$: a broad range of conditions at which thermochemical equilibrium between the surface and the lower atmosphere, and in the lower atmosphere itself, may occur.
The temperature at the highest surface altitude on Venus is $650\,\text{K}$; \citet{liggins23} find that volcanically-derived atmospheres can only achieve thermochemical equilibrium in the gas phase at $T\gtrsim700\,\text{K}$.
The upper bound of $1700\,\text{K}$ is roughly the temperature at which quartz melts \citep{chase86}: in GGchem this was found to be the major mineral which melts at the lowest temperature.
In fact, melting is likely to become important at lower temperatures than this, due to the stabilisation of eutectic melts.
Results above around $1500\,\text{K}$ should therefore be treated as illustrative only.
We elaborate on this point in Section~\ref{ggchemlimitations}.

In these calculations we again use the estimated elemental composition of Venus as given in Appendix \ref{app:elements}.

Abundances of several minerals are shown in the top row of Figure~\ref{fig:grids}.
Each plot shows an arc-shaped boundary, from $(p_0\,/\,\text{bar},\,T_0\,/\,\text{K})\approx(10^{-1},1200)$ to $(10^3, 2000)$, across which the mineral abundances change sharply.
This is particularly stark for \ce{CaSO4}, diopside (\ce{CaMgSi2O6}), and spinel (\ce{MgAl2O4}), each of which disappears completely on one side of the boundary\footnote{
Spinel is also lost isothermally below $\sim1200\,\text{K}$, though this is likely to be an example of what \citet{woitke18} refer to as a type-2 transition, where one mineral is suddenly switched off.
This behaviour would not be exhibited in a natural system.
}

The equilibrium VMRs of \ce{CO2}, \ce{O2}, \ce{SO2}, and \ce{SO3} are shown in the bottom row of Figure~\ref{fig:grids}.
There is a clear transition in the atmospheric chemistry, across a boundary which roughly coincides with the boundaries seen in the mineral abundances in the top row.
Whereas on one side the atmospheres are \ce{CO2}-dominated, on the other they consist of a mixture of \ce{SO2} and \ce{O2}, with \ce{SO3} constituting a minor component near to the boundary.

Mass balance investigations show that the boundary on the top row of Figure~\ref{fig:grids} are conceptually due to the two reactions:
\begin{equation} \label{eq:diopside}
\ce{
CaMgSi2O6 + SO2 + 1/2O2
<-> CaSO4 + MgSiO3 + SiO2
};
\end{equation}
\begin{multline} \label{eq:anorthite}
\ce{
CaAl2Si2O8 + SO2 + 1/2O2 + MgSiO3\\
<-> CaSO4 + MgAl2O4 + 3SiO2
};
\end{multline}
each of which involves the oxidation of \ce{SO2} into \ce{CaSO4} at the expense of a calcium-bearing mineral and \ce{O2}. These reactions occur at slightly different pressure--temperature conditions, causing the formation of \ce{CaSO4} to occur in two stages as the pressure is raised (or as the temperature is lowered).

The change in the VMR of \ce{CO2} across the boundary is not related to the stability of any carbon-bearing mineral\footnote{
There is in fact another boundary at much higher pressure and lower temperature at which \ce{CO2} is released by the decomposition of \ce{MgCO3}, but we do not consider it of further relevance to the Venus-like planet scenarios.
}, unlike with \ce{SO2} and \ce{O2}:
at least 99.4 per cent of the carbon atoms are contained within \ce{CO2} on both sides of the boundary.
\ce{CO2} is therefore present across almost the entire grid.
On the lower-pressure side of the boundary it is vastly outnumbered by \ce{SO2} and \ce{O2}, which dominate the atmosphere under these conditions.
At higher pressures, reactions~\ref{eq:diopside} and \ref{eq:anorthite} sequester \ce{SO2} and \ce{O2} into \ce{CaSO4}, leaving \ce{CO2} as the dominant component.

One important note about these simulations is that as \textsc{GGchem} is run at fixed p and T.  The consequence of this is that reactions sequestering gas phase species lead to a multiplying up of the actual number density of every species in proportion so that the atmosphere maintains the requisite pressure.
Therefore, although the element ratios are preserved between different points on the grid, each point has a different system mass.
We therefore perform mass balance analyses by looking at the \textit{fraction} of the atoms of a given element that are contained within different species.
For example, on the higher-pressure side of the boundary, almost all of the sulphur atoms are in \ce{CaSO4}; on the lower-pressure side almost all are in \ce{SO2} and \ce{SO3}.

\subsection{Transmission spectra across mineral-gas regimes}

\begin{figure*}
\centering
\includegraphics[width=\textwidth]{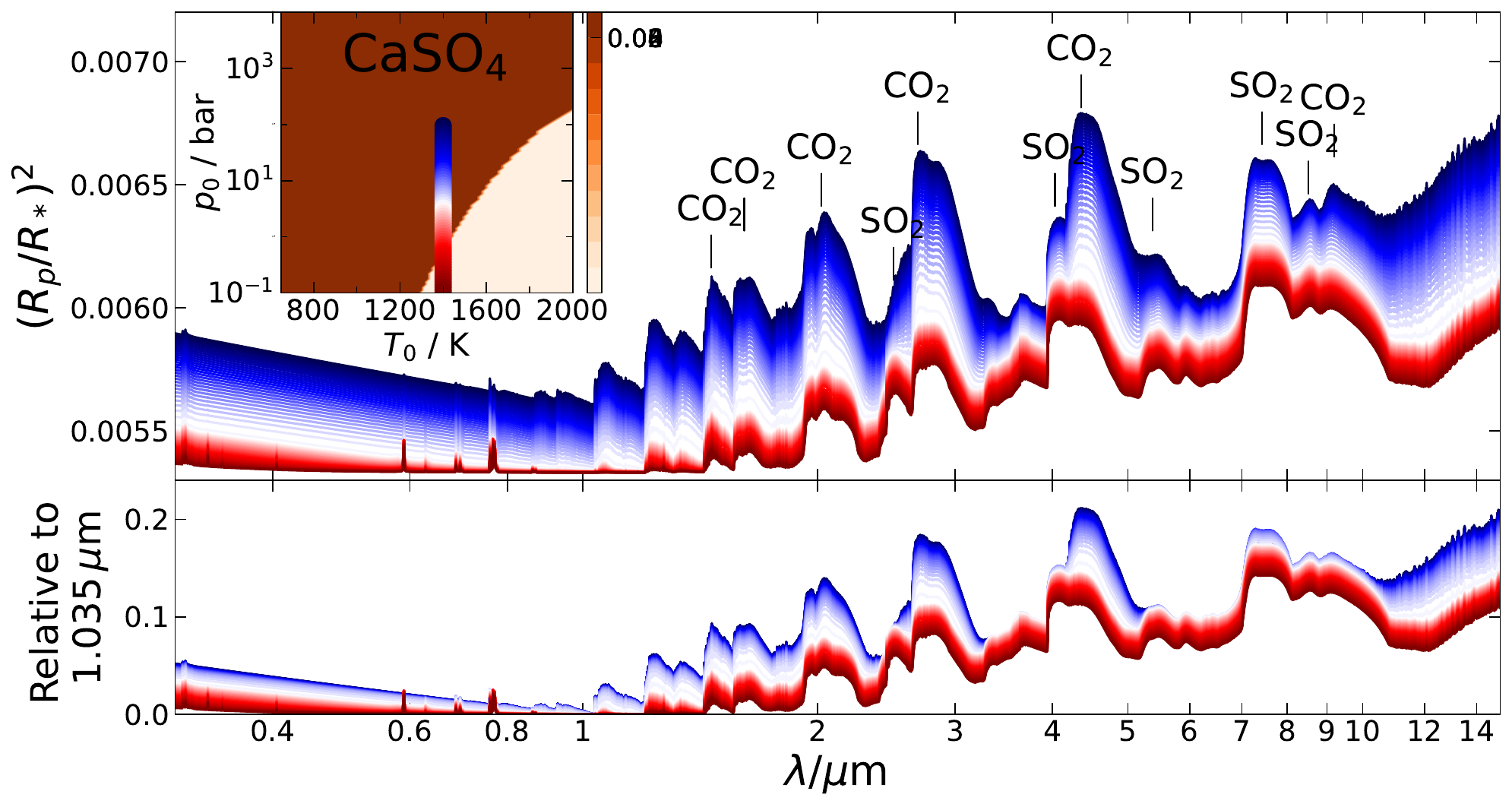}
\caption{
Idealized transmission spectra across the boundary between atmospheric regimes.
The inset shows an isothermal transect through $p_0$--$T_0$ space, at $T_0=1400\,\text{K}$, and crossing between the two surface/atmospheric regimes.
Assuming planets of Venus' mass and radius, orbiting a TRAPPIST-1-like star, transmission spectra are shown for homogeneous isothermal atmospheres, whose surface conditions are colour-coded to the inset.
The transit radius increases almost monotonically as the surface pressure is increased, due to the reduction in the mean molecular weight on transitioning from an atmosphere dominated by SO$_2$ ($\mu\approx64$) to one dominated by CO$_2$ ($\mu\approx44$).
The lower figure shows the relative deviation in the transit radius from that at $1.035\,\mu\text{m}$, and emphasizes the very different spectral shapes, particularly between $2\,\mu\text{m}$ and $6\,\mu\text{m}$.}
\label{fig:transect-homoiso}
\end{figure*}

We have seen above that the compositions of the high and low temperature (low and high pressure, respectively) atmospheric regimes are starkly different.
It follows that transmission observations of the atmosphere of a Venus-composition exoplanet could identify whether the atmosphere is \ce{CO2}- or \ce{SO2}-\ce{O2}-dominated, and hence whether the surface pressure--temperature conditions are above or below the boundary.
As the mineralogy is also, to first order, dependent on which side of the boundary the surface conditions are, such observations would then also provide unprecedented information about the planet's mineralogy.

Using the radiative transfer package \textsc{petitRADTRANS} \citep{petitRADTRANS}, and the comprehensive line lists provided by HITRAN \citep{hitran} and ExoMol \citep{exomol}, we generated idealized transmission spectra for a series of hypothetical Venus-like exoplanets whose surface conditions lie on a transect straddling the identified gas--mineral boundary (Figure~\ref{fig:transect-homoiso}).
These spectra were calculated for a planet orbiting a TRAPPIST-1-like star.
We emphasize that these transmission spectra are intended only to be illustrative, as they have been generated under the unrealistic assumption of a homogeneous and isothermal atmosphere.

However, these spectra do serve to illustrate that atmospheres on either side of this boundary are likely to be distinguishable by transmission spectroscopy, and that observations could therefore identify the surface pressure--temperature regime -- and hence mineralogical regime -- of a warm rocky exoplanet.
For example, spectral features due to \ce{SO2} at $2.5\,\micron$, $4.0\,\micron$, and $5.4\,\micron$ are prominent in one regime but muted in the other, where the \ce{CO2} feature at $4.4\,\micron$ is more prominent.

Atmospheric inhomogeneity, lapse rate, photochemistry and other non-equilibrium processes all undoubtedly influence the chemistry of the upper atmosphere \citep[to which infrared transmission spectra are primarily sensitive; e.g.][]{kitzmann18} and further work would be necessary to determine the best ways to distinguish these atmospheric regimes in practice.
None the less, we suggest that the chemistries of these atmospheres are sufficiently distinct that accounting for these complications would not erase the identification of the two regimes in real observations.
This is discussed further in Section~\ref{caveats}.

\section{Role of a planet's elemental composition in setting atmospheric regime}
\label{composition}

The correlation described above between atmospheric and mineralogical regimes has been calculated based on the composition of the crust being equal to that inferred for Venus \citep{rimmer21}.
While short-period (warm) exoplanets may have surface pressure--temperature conditions conducive to achieving gas-mineral thermochemical equilibrium, the existence of a large population of exoplanets with identical crustal compositions to Venus is unlikely.

Estimating the bulk composition of an exoplanet is challenging in practice (\citealt{bond10, elser12}), though constraints may be placed based on the refractory composition of the host star, as obtainable by spectroscopy \citep[e.g.,][]{hinkel14,thiabaud15}.
The composition of the crust can then be estimated from that of the bulk planetary mantle using thermodynamics codes which simulate magmatic processes, such as \textsc{MELTS} \citep{ghiorso95}.
Although the link between the stellar and planetary crustal compositions is contingent on unknown properties of the system, previous work may none the less enable a rough estimation of the relative elemental abundances of refractory elements in an exoplanet's crust.  From this variation, we can then evaluate the robustness of the mineral-atmosphere regimes of Figure~\ref{fig:grids} and therefore understand how well-constrained the composition of a planet needs to be for its surface mineralogy to be inferable from atmospheric observations.

We address this surface--atmosphere link in the context of changing crustal composition in this section.
Starting from Venus' composition, we vary the abundances of several important elements in turn, and assess the sensitivity of the correlation between the mineralogy and the atmospheric chemistry to the underlying crustal composition.
We focus on the same isothermal transect as in Figure~\ref{fig:transect-homoiso} ($1400\,\text{K}$, running from $10^{-1}\,\text{bar}$ to approximately $10^{2}\,\text{bar}$) for simplicity and ease of visualisation.
The compositional ranges we have investigated for refractory elements are motivated by the variation among stars in the Hypatia catalog \citep{hinkel14} and solar system geology.  These ranges are likely representative of diverse planetary crusts.

\subsection{Calcium}

In our calculations the transformation from an \ce{SO2}-\ce{O2}-dominated atmosphere to a \ce{CO2}-dominated atmosphere is driven by the incorporation of \ce{SO2} and \ce{O2} into \ce{CaSO4}.
The availability of calcium is thus likely to impact the nature of the boundary between the two regimes.
Na\"{i}vely increasing the abundance of \ce{Ca} on its own would severely disrupt the redox state of the system, and lead to unrealistic highly-oxidized or -reduced systems: any added Ca would consume available oxygen to produce CaO, and Ca removed would leave behind O.
We avoid this issue by adding/subtracting stoichiometric CaO, effectively altering the CaO weight percentage (wt\%) of the crust.
The relative abundances of all other elements are kept constant.

\begin{figure}
\centering
\includegraphics[width=\columnwidth]{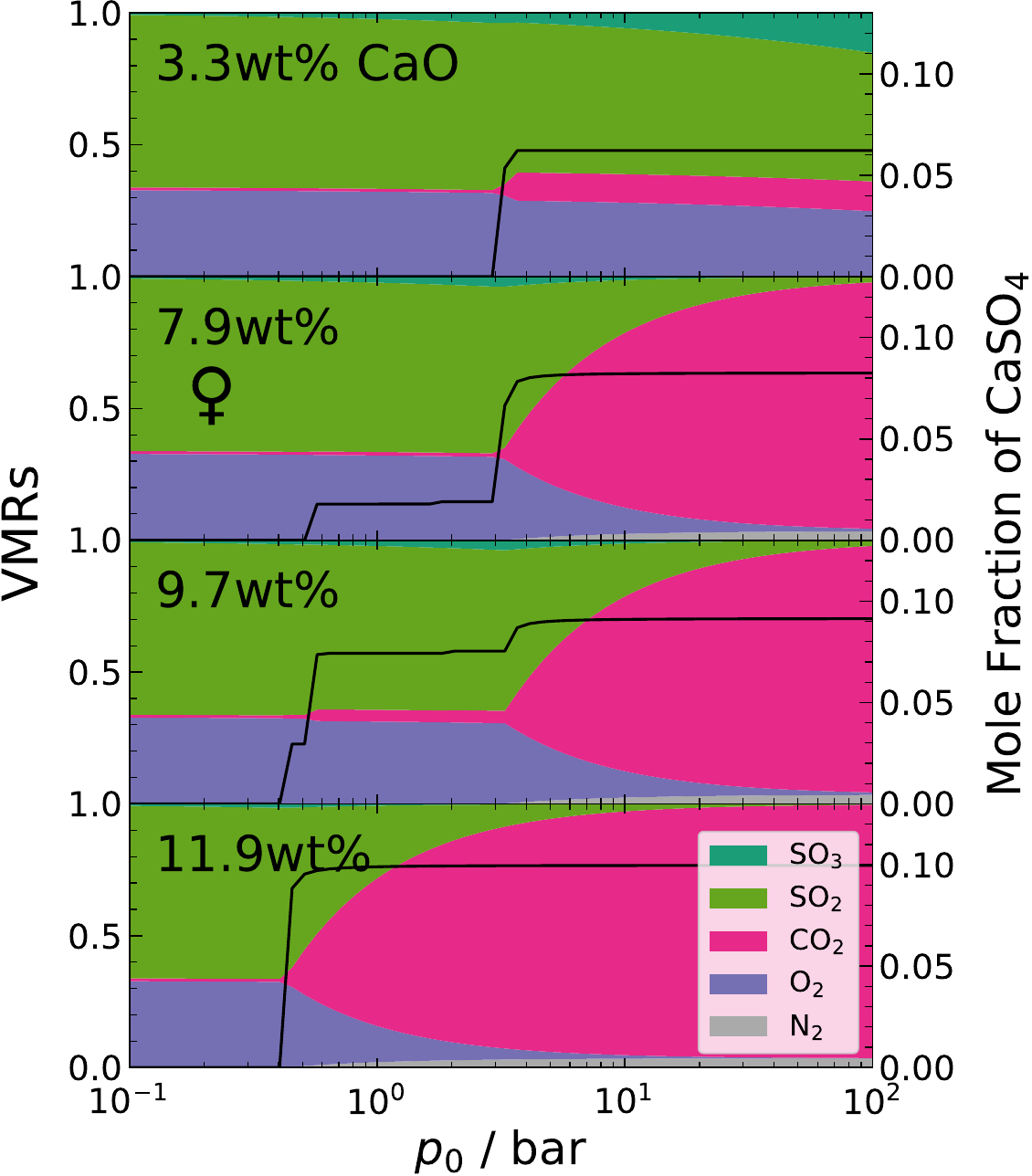}
\caption{
Equilibrium atmospheric compositions for different calcium abundances at $1400\,\text{K}$.
The black lines show the abundances of \ce{CaSO4} as a function of pressure.
The second panel used Venus' elemental composition.
Depending on the CaO wt\%, the sulphidation of the calcium bearing minerals -- and hence the change in atmospheric composition -- occurs at different pressures, in different stages, and to different extents.
}
\label{fig:Catransect}
\end{figure}

Figure~\ref{fig:Catransect} shows the equilibrium atmospheric compositions and the mole fraction of \ce{CaSO4} for four calcium abundances, expressed as the CaO wt\% of the solids.
We see that in each panel -- independent of the available calcium -- the atmosphere is \ce{SO2}-\ce{O2}-dominated at pressures lower than about $p_0=0.4\,\text{bar}$.
At higher pressures, \ce{SO2} and \ce{O2} are incorporated into \ce{CaSO4}, leaving \ce{CO2} to prevail as a significant component of the atmosphere.
However, the pressure range over which this occurs does depend on the CaO wt\%, as does the eventual contribution of \ce{CO2} to the atmospheres at high pressure.

For similar calcium abundances to Venus, mass balance investigations show that reaction~\ref{eq:diopside} occurs at $p_0=0.5\,\text{bar}$, then reaction~\ref{eq:anorthite} at $p_0=4\,\text{bar}$.
(These pressures given for the above reactions only apply at $T_0=1400\,\text{K}$; Figure~\ref{fig:grids} shows that at higher temperatures these reactions occur at higher pressures.)

The second panel of Fig.~\ref{fig:Catransect} shows that for Venus' composition, reaction~\ref{eq:diopside} (where diopside acts as a sink for atmospheric sulphur) has little effect on the atmospheric chemistry, with the VMR of \ce{CO2} increasing only slightly from 0.011 to 0.013.
There is then a range of pressures for which a small amount of \ce{CaSO4} is present yet atmospheric \ce{CO2} is still very low.
For higher CaO wt\%, however, reaction~\ref{eq:diopside} desulphidates the atmosphere to such an extent that \ce{CO2} attains a significant VMR at $p_0 = 0.5\,\text{bar}$ and above.
The more calcium is available, the more \ce{CaMgSi2O6} is present in the low-pressure mineralogy compared to \ce{CaAl2Si2O8}, enabling more \ce{SO2} to be incorporated into the crust at this lower pressure, rather than the higher pressure at which \ce{CaSO4} is formed from \ce{CaAl2Si2O8}.

By 11.9$\,$wt\% CaO, there is enough \ce{CaMgSi2O6} to react with all of the atmospheric \ce{SO2} at low pressure.
This proceeds through a different reaction at a slightly lower pressure of $p_0=0.4\,\text{bar}$, facilitated by the conversion of \ce{Mg2SiO4} (forsterite, not present for lower calcium abundances) to \ce{MgSiO3}:
\begin{equation} \label{eq:diopsidelp}
\ce{
CaMgSi2O6 + SO2 + 1/2O2 + Mg2SiO4
-> CaSO4 + 3 MgSiO3
}.
\end{equation}

For Ca-poor crusts, there is little or no \ce{CaMgSi2O6} present at low pressure, and the \ce{SO2} can only be absorbed by \ce{CaAl2Si2O8} at $4\,\text{bar}$.
At 3.3$\,$wt\% CaO, there is not even sufficient \ce{CaAl2Si2O8} to absorb all of the \ce{SO2}, and some sulphur remains in the atmosphere even at high pressure.
Enough is sequestered in the crust, however, that \ce{CO2} is able to constitute a mixing ratio of order 0.1 in the high-pressure atmospheres, as shown in the top panel of Figure~\ref{fig:Catransect}.

Although the correlation between the mineralogy and the atmospheric chemistry is more nuanced when the amount of calcium is uncertain, some results remain valid.
For the vast majority of compositions and conditions, the presence of \ce{CaSO4} is associated with \ce{CO2} being present in the atmosphere; the only exceptions being where the surface pressure is between $0.5$ and $4\,\text{bar}$ and the CaO wt\% is between about 4 and 9 per cent.
Being a highly radiatively active gas, it is likely that \ce{CO2} would be detectable by transmission spectroscopy even if its VMR is small, enabling the inference of surface \ce{CaSO4} for a range of calcium abundances.

\subsection{Aluminium}

For Venus' elemental composition, the mineral which is responsible for most of the high-pressure sulphidation of the crust is \ce{CaAl2Si2O8}.
The crust's aluminium content is therefore also likely to be an important factor in determining its effect at thermochemical equilibrium on atmospheric chemistry.
Again, to preserve the oxidation state we add 1.5 O atoms for every Al atom, effectively adding \ce{Al2O3} to the composition.

Increasing the amount of Al has a very similar effect to \textit{decreasing} the amount of Ca.
Both changes to the crustal composition increase the amount of \ce{CaAl2Si2O8} relative to \ce{CaMgSi2O6} at low pressure, which causes more of the \ce{SO2} to be absorbed at $4\,\text{bar}$ and less at $0.5\,\text{bar}$.
Surfaces which have a high \ce{Al2O3} weight percentage therefore show an abrupt boundary at around $4\,\text{bar}$ where \ce{CaAl2Si2O8} absorbs all the atmospheric sulphur to form CaSO$_4$ (by reaction~\ref{eq:anorthite});
this occurs when the \ce{Al2O3} is $>18\,\text{wt\%}$ (cf.\ Venus: $(16\pm1.8)\,\text{wt\%}$,  \citealt{surkov86}).
Conversely, surfaces with very \textit{low} \ce{Al2O3} wt\% ($\lesssim 2.5\,\text{wt\%}$) have enough of their calcium present as \ce{CaMgSi2O6} to absorb all the atmospheric sulphur by reaction~\ref{eq:diopside}, giving instead a sharp boundary at $0.5\,\text{bar}$.

\subsection{Magnesium}

\begin{figure}
\centering
\includegraphics[width=\columnwidth]{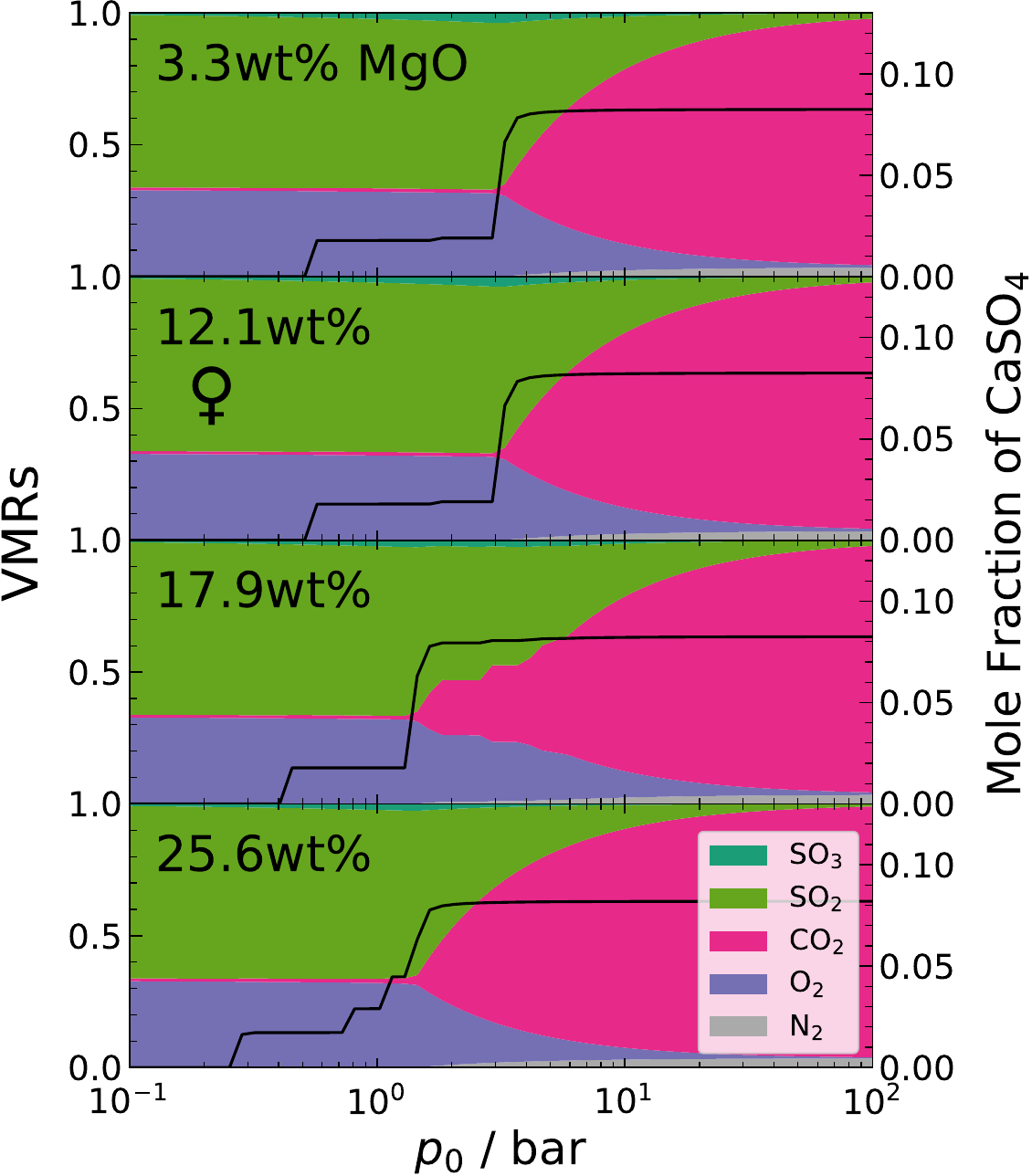}
\caption{
Equilibrium atmospheric compositions for different magnesium abundances at $1400\,\text{K}$.
Increasing the amount of \ce{MgO} in the crust significantly complicates the sulphidation of the crust, which proceeds in several steps.
Decreasing the amount of \ce{MgO}, even very significantly, has little effect.
}
\label{fig:Mgtransect}
\end{figure}

Figure~\ref{fig:Mgtransect} shows the atmospheric chemistry and \ce{CaSO4} abundance at the boundary when the amount of MgO is increased from Venus' level.
Reducing the amount of \ce{MgO} compared to Venus has very little effect.
This could be due to magnesium's only influence on the chemistry being through \ce{CaMgSi2O6} and \ce{Mg2SiO4}, the latter of which is absent for magnesium contents equal to or less than Venus'.

With a higher MgO content, some magnesium \textit{is} found as \ce{Mg2SiO4} at low pressure.
This enables some \ce{SO2} to be absorbed at $0.4\,\text{bar}$, by reaction~\ref{eq:diopsidelp}, as was the case for very CaO-rich crusts.
If MgO is $\gtrsim$13$\,\text{wt\%}$, there will still be some \ce{Mg2SiO4} remaining after this reaction.
This then becomes unstable at around $1.5\,\text{bar}$, forming \ce{MgAl2O4} and \ce{MgSiO3} by the reaction:
\begin{multline}
\ce{
3Mg2SiO4 + CaAl2Si2O8 + SO2 + 1/2O2\\
-> MgAl2O4 + 5MgSiO3 + CaSO4
},
\end{multline}
leading to \ce{SO2} being absorbed from the atmosphere at another intermediate pressure.
For $\geq$17.9$\,\text{wt\%}$ MgO, the overall picture is complicated by the presence of more exotic phases such as \ce{MgFe2O4} (magnesioferrite) and \ce{Mn3Al2Si3O12} (spessartine), and their auxiliary interactions with the sulphur chemistry.
These species cause the emergence of \ce{CO2} in the atmosphere to proceed in several steps, likewise the appearance of \ce{CaSO4} in the crust.

The interference of these additional Mg-bearing minerals with the sulphur chemistry would complicate mineralogical inferences in practice, particularly if an exoplanet's surface conditions and composition are poorly-constrained initially.
However, the subtleties which emerge are confined within a relatively narrow region of $p_0$--$T_0$ space around the boundary, and if other observations suggest that a planet's surface conditions are far away from the boundary then such inferences would remain feasible.

\subsection{Oxygen}

As the most abundant metal in the Universe -- and the largest constituent of terrestrial planets -- variation in the amount of available oxygen is likely.
It has long been suggested that Venus itself may have once had abundant surface water that was subsequently lost, with a runaway greenhouse effect followed by photolysis of \ce{H2O} and hydrogen escape \citep[e.g.,][]{kasting84}, resulting in a significant release of \ce{O2} into the atmosphere.
\ce{O2} may also be released due to photolysis of \ce{CO2} \citep{wong19}.
Whatever its source, liberated oxygen may oxidize surface minerals before accumulating in the atmosphere, leading to an abiotic buildup of \ce{O2} \citep[e.g.,][]{luger15}.
It is therefore prudent to explore the effect of altering the amount of available oxygen to the system.

For each oxygen abundance, we again run \textsc{GGchem} along the same isothermal track in $p_0$--$T_0$ space.
The oxygen abundance for a given run is quantified by the fugacity of \ce{O2} at the pressure of $95\,\text{bar}$, $\fugo$.
$f_{\ce{O2}}$ is a useful quantifier of how oxidising the system is and the oxidation state of the surface \citep{fegley97}, and a pressure of $95\,\text{bar}$ is chosen to aid comparison with Venus.
At $1400\,\text{K}$, this pressure is in the regime where the equilibrium atmospheres are \ce{CO2}-dominated and \ce{O2}-poor, as with Venus.
Venus' surface conditions are however at much lower temperatures, far away from the boundary across which the atmosphere becomes \ce{SO2}-dominated.
The $\fugo$ values given below, being measured at surface conditions much closer to the boundary, are much higher than the $f_{\ce{O2}}$ on Venus' surface; \citet{fegley97} give $\log f_{\ce{O2}}$ between $-21.7$ and $-20.0$ at $0\,\text{km}$.

\begin{figure}
\centering
\includegraphics[width=\columnwidth]{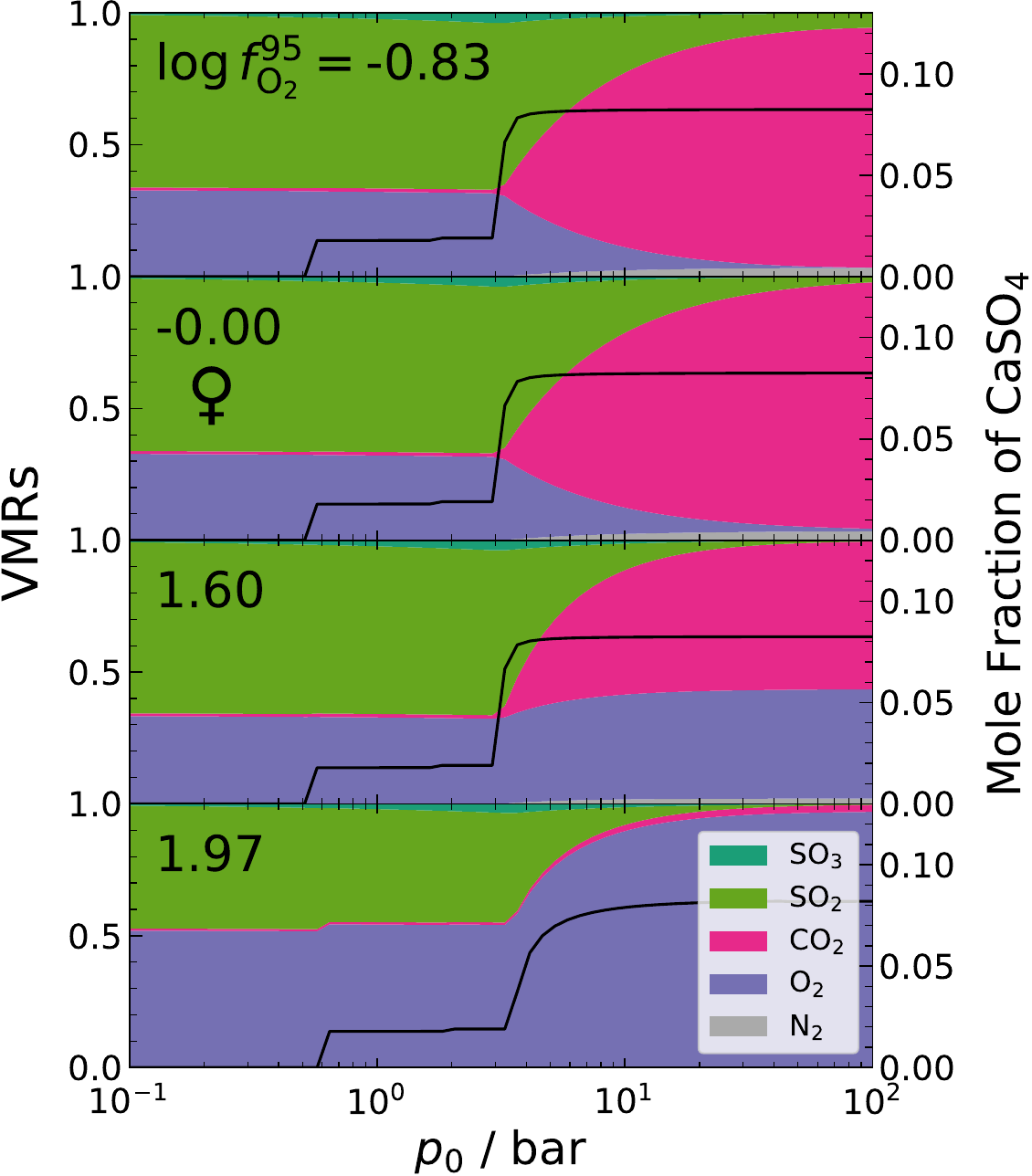}
\caption{
Equilibrium atmospheric compositions for different oxygen abundances at $1400\,\text{K}$.
By coincidence, Venus' elemental composition gives an equilibrium $f_{\ce{O2}}$ very close to $1\,\text{bar}$ at $p_0=95\,\text{bar}$ and $T_0=1400\,\text{K}$, hence $\log\fugo\approx0$ for this composition.
Changing the amount of available oxygen has little effect on the formation of \ce{CaSO4}, any excess oxygen simply going into atmospheric \ce{O2}.
}
\label{fig:Otransect}
\end{figure}

The atmospheric composition and abundance of \ce{CaSO4} for varying oxygen abundances are shown in Figure~\ref{fig:Otransect}.

Any oxygen which is added relative to Venus' composition has no noticeable effect on the sulphur chemistry, instead simply increasing the amount of \ce{O2} in an already fully-oxidized system.
When enough oxygen is added that $\log\fugo>1.5$, there is so much atmospheric \ce{O2} that when \ce{SO2} and \ce{O2} are incorporated into crustal \ce{CaSO4}, the loss of atmospheric \ce{SO2} actually causes the VMR of \ce{O2} to increase.
In such oxygen-rich cases, the high-pressure atmospheres are dominated by an increasingly \ce{O2}-rich mixture of \ce{CO2} and \ce{O2}.

Although these atmospheres are very different from those emerging from Venus' composition, the observational distinguishability of the two regimes is likely to remain.
As a homonuclear diatomic molecule, \ce{O2} has very low opacity at infrared wavelengths, and hence would only have a direct effect on exoplanet transmission spectra at visible and UV wavelengths.
The more IR-active species \ce{CO2} and \ce{SO2} would have a reduced impact on the spectrum compared with the low-oxygen cases, as their VMRs are drowned out by the dominant \ce{O2}, but with much higher opacities they would likely still be diagnostic of the surface state.
Even atmospheres flooded with \ce{O2} are therefore expected to retain the surface--atmosphere correlation we have described.

\subsection{Sulphur}

\begin{figure}
\centering
\includegraphics[width=\columnwidth]{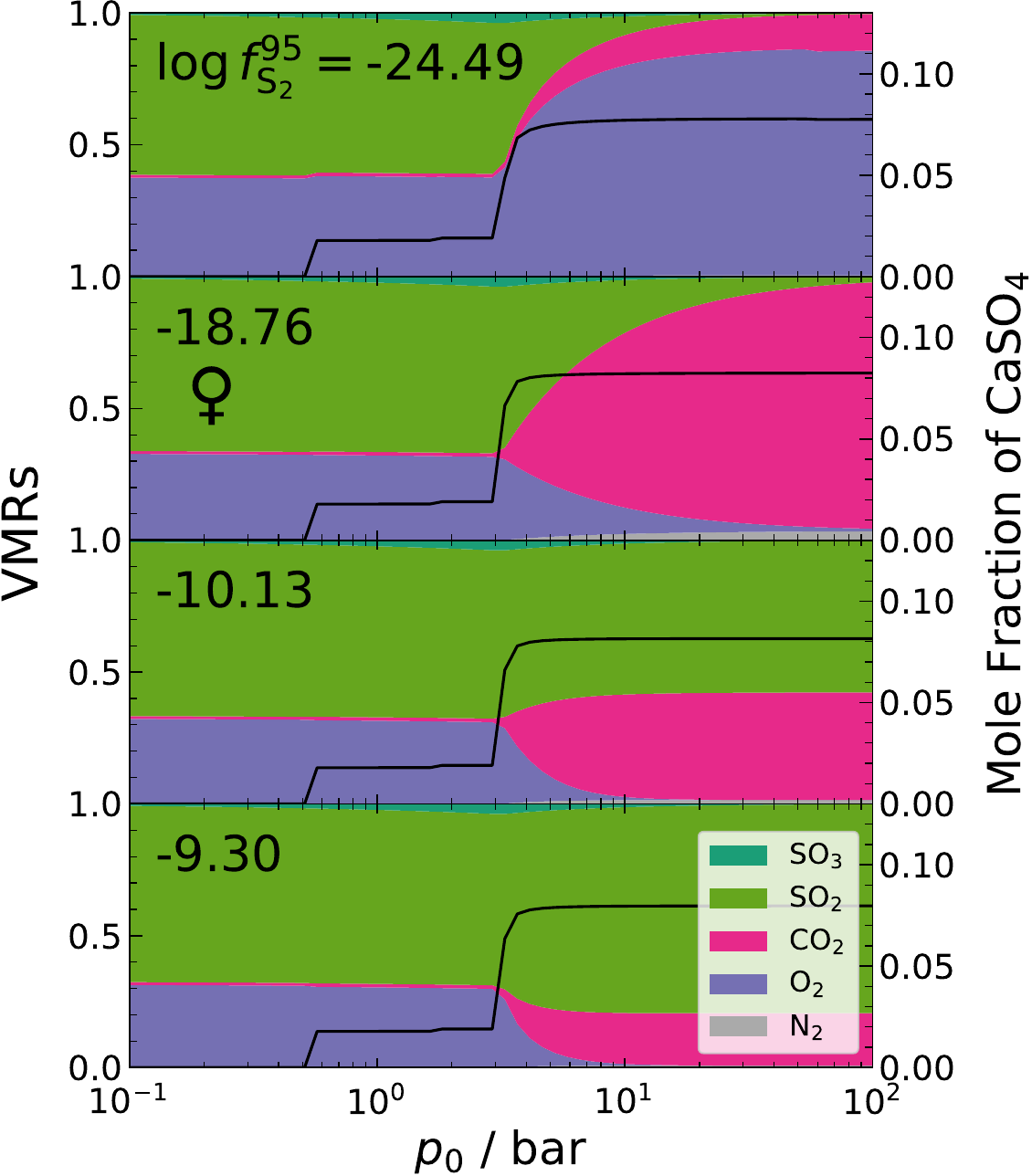}
\caption{
Equilibrium atmospheric compositions for different sulphur abundances at $1400\,\text{K}$.
Altering the amount of sulphur has no noticeable effect on the sulphidation process, simply changing the balance between \ce{SO2} and \ce{O2} in the high-pressure atmospheres.
}
\label{fig:Stransect}
\end{figure}

Sulphur's transference from atmospheric \ce{SO2} to \ce{CaSO4} forms the dividing line between the two mineral-atmosphere regimes identified above.  It is therefore important to understand the effect of changing its abundance.
By analogy with the previous subsection, we measure the amount of available sulphur through the parameter $f_{\ce{S2}}^{95}$, which on the surface of Venus is thought to be between $10^{-4.2}$ and $10^{-6}\,\text{bar}$ \citep{zolotov18}.
Again, Venus' surface conditions are very different to those near the regime boundary we are investigating, so here we find very different fugacities.

The effects of altering the amount of available sulphur are shown in Figure~\ref{fig:Stransect}.
As with oxygen, changing the sulphur has little effect on the absorption of sulphur into the crust, and the VMR of \ce{CO2} increases when \ce{CaSO4} sequesters the atmospheric \ce{SO2} and \ce{O2}.
The only major effect is the balance between \ce{SO2} and \ce{O2} in the atmosphere.
Reducing the amount of sulphur has a similar effect to increasing the amount of oxygen.
The atmosphere becomes dominated by \ce{CO2} and \ce{O2} when \ce{SO2} is removed into the crust, as shown in the top panel of Figure~\ref{fig:Stransect}.
Increasing the amount of sulphur simply increases the VMR of \ce{SO2} in the high-pressure atmospheres.
When $\log \fugs > -10$, the atmosphere is so \ce{SO2}-rich that when \ce{SO2} and \ce{O2} form \ce{CaSO4} the loss of \ce{O2} actually causes the \ce{SO2} VMR to increase.

Increasing the amount of sulphur leads to the buildup of a radiatively-active gas (\ce{SO2}) superceding the \ce{CO2} at high pressure.
In fact, the bottom panel of Figure~\ref{fig:Stransect} shows that there is actually a higher VMR of \ce{SO2} at high pressure than at low pressure.
When observing such atmospheres, the presence of \ce{SO2} is no longer indicative of a lack of surface \ce{CaSO4}, or the presence of \ce{CaAl2Si2O8} or \ce{CaMgSi2O6}.
\ce{CO2} maintains its association with the sulphidated mineralogy, though its reduced VMR may hinder its detectability if the exoplanet being observed is more sulphur-rich.

\section{Discussion}
\label{discussion}

\subsection{Implications}

We have demonstrated that the atmospheres and surfaces of warm rocky exoplanets, when coupled sufficiently strongly by thermochemical equilibrium, may enable observational inferences on the surface mineralogy of such worlds.
Facilities already in operation are more than capable of distinguishing between atmospheres that are dominated by \ce{CO2} and those that are dominated by \ce{SO2}.
Observations using such facilities could therefore determine whether a planet's exterior sulphur reserves are present in its atmosphere or in its crust; we have shown that the intermediate region of surface conditions -- where some sulphur is in both -- is rather narrow for most compositions.

Knowledge of where the sulphur is also gives a loose constraint on the surface pressure and temperature, information which is difficult to otherwise constrain.
Furthermore, knowing whether or not the crust is sulphidated informs the mineralogy: whether calcium is present as \ce{CaSO4}, or as \ce{CaAl2Si2O8}; whether magnesium is present as \ce{MgAl2O4}, or as \ce{CaMgSi2O6}.
Inferences on exoplanet mineralogies are as yet unprecedented in astronomy, and represent the natural next step in our ever-more-detailed understanding of the exoplanet population.

Future work would employ the principle of surface--atmosphere equilibrium as a useful lower boundary condition, fixing the atmospheric composition at low altitudes.
A natural next step would be to couple the equilibrium to a 1D atmospheric model.

\subsection{Caveats and future directions to modelling surface--atmosphere interaction} \label{caveats}

\subsubsection{Chemical Gradients and Non-Equilibrium Processes}

The theoretical study of surface--atmosphere equilibrium can of course only yield information about the chemistry at the base of the atmosphere.
Observations of exoplanetary atmospheres however probe only the uppermost layers of the atmosphere \citep[$p\lesssim 10^{-2}\,\text{bar}$, e.g.,][]{kitzmann18}.
As such, to be able to practically exploit the correlation that we have identified between the surface and atmospheric regimes, we require the chemical distinctiveness of the atmospheric regimes to carry through to higher altitudes/lower pressures.  This does not necessarily require the exact chemistry to be homogeneous throughout the atmosphere, but rather that surface regimes none the less produce distinct chemistry at higher altitudes.
For example, atmospheres which are \ce{SO2}-rich at the surface may instead contain \ce{SO3}, \ce{SO}, \ce{COS}, or \ce{H2S} at higher altitudes, simply due to the relevant sulphur equilibria shifting at lower temperatures and pressures.
Though not the direct product of the reactions occurring at the surface, the presence/absence of these species is still capable of indicating whether sulphur is absent/present from the surface mineralogy, and would likely be detectable by transmission spectroscopy if present at high enough VMRs.

Photochemical processes may facilitate the above conversions between sulphur species, or alternatively may oxidize \ce{SO2} to \ce{H2SO4} droplets, as occurs on Venus \citep[e.g.,][]{krasnopolsky94}.
This process removes \ce{SO2} from Venus' atmosphere efficiently, with the VMR of this species of order $10^4$ lower above the clouds compared to below them \citep{mills07}.
This may limit the distinguishability of the two regimes: if all of the sulphur in the \ce{SO2}-rich base atmospheres is converted to species which all condense out into clouds and precipitation at higher altitudes, the atmosphere would appear to be sulphur-poor, and one might erroneously infer the presence of \ce{CaSO4} on the surface, where in fact the sulphur is all located above the surface but below the clouds.
However, at least for M-dwarf host stars, \citet{jordan21} find that their lower UV-flux ensures that \ce{SO2} and other sulphur species survive above the cloud layer of Venus-like planets.
For these objects, surface-atmosphere coupling may still be visible in the chemistry of the high-altitude atmosphere.

Vertical mixing can also strongly impact the observable atmospheric chemistry \citep[e.g.,][]{prinn77}.
Although the detailed effects of this phenomenon are beyond the scope of this work, substantial vertical mixing would advect the equilibrium chemistry of the surface up to higher altitudes where it is observable.
This tendency to homogenize the atmosphere would act to preserve the distinguishability of the two atmospheric classes (S-rich and S-poor) into the upper atmospheres.  Vertical mixing therefore enables transmission spectroscopic observations to more readily identify the regime of the surface mineralogy.

\subsubsection{3D planetary effects}

The models we have considered are essentially 0-dimensional and do not account for 3D effects that could alter the atmospheric structure and composition.
Many known exoplanets are tidally locked, with drastic differences in the conditions on the dayside and nightside of the planet.
Different regions of a given planet may therefore occupy different parts of $p_0$--$T_0$ space, perhaps even crossing the boundary we have found.
Indeed, perhaps the day--night temperature difference would be so large that surface--atmosphere equilibrium would only be in effect at certain regions of the surface.
Counteracting this is atmospheric superrotation, as occurs on Venus \citep[e.g.,][]{read18}, which would mitigate the effects of tidal locking:
strong zonal winds would longitudinally homogenize the surface conditions, preventing the strong temperature differential from being communicated to the surface.

Planet hypsometry could have a similar effect: variations in surface altitude (and hence pressure and temperature) may lead to a different mineral assemblages and atmospheric compositions at different altitudes.
Indeed, we showed in Section~\ref{methods} that this is the case on Venus: above a certain altitude the mineralogy may change slightly as \ce{FeS2} becomes stable.
However, this is also likely to be a second-order effect:
the equilibrium \textit{atmosphere} does not drastically change at the altitude where \ce{FeS2} becomes stable; indeed, only a small amount of \ce{FeS2} is formed.
Conversely, this means that \ce{FeS2} would not have sufficient impact on the atmospheric composition for its presence to be inferable from transmission observations of Venus.
We therefore stress that surface--atmosphere equilibrium would not permit the inference of altitudinal variations in the mineralogy on a given exoplanet.

\subsubsection{Planetary compositional variability}

Section~\ref{composition} investigated the robustness of the mineralogy--atmosphere correlation to variations in the underlying elemental composition, finding that the correlation was maintained for the majority of cases, with certain caveats.
We have not, however, explored the full extent of the vast multi-dimensional space of elemental compositions; we have merely looked at the effect of varying certain elements individually.
It is possible that crusts which are both, say, more \ce{Ca}- \textit{and} \ce{Mg}-rich than Venus' would exhibit different behaviour.
The purpose of this work is simply to demonstrate the principle of relating atmospheric and mineralogical compositions via thermochemical equilibrium, rather than to provide an accurate road map of how the two are related in all possible cases.

The closer an exoplanet's elemental composition is to Venus', the more directly applicable our results are, but it would not be difficult to apply the principles outlined here to planets of differing composition.
Our results here demonstrate that, for roughly Venus-composition planets, a boundary exists in $p_0$--$T_0$ space between both atmospheric and mineralogical regimes.
Perhaps in other cases it would be different minerals, or different atmospheric species, that would relate to one another to an observably distinguishable extent.
We see no reason why Venus' composition would be special in exhibiting such a boundary between atmospheric and mineralogical regimes.

\subsubsection{Sharpness of the regime boundary}

The boundary between the two atmospheric/mineralogical regimes is not completely sharp.
For most compositions, there is an intermediate regime of surface conditions for which the crust is only partly sulphidated, with some sulphur present in the crust as \ce{CaSO4} and some in the atmosphere as \ce{SO2} or \ce{SO3}.
At $1400\,\text{K}$, as examined above, this corresponds to surface pressures between around $0.4$ and $4\,\text{bar}$.
If the surface conditions happen to lie in this regime, inferences on the mineralogy based on the atmospheric chemistry would be more difficult, as discussed at various points in the previous section.
None the less, we are reassured by the narrow width of this intermediate region as shown in Figure~\ref{fig:grids} for Venus' composition.
Although the surface conditions of rocky exoplanets are unlikely to be uniformly distributed across the ranges examined, we see no reason why a large proportion should lie within this strip.

\subsubsection{Equilibrium chemistry code limitations}
\label{ggchemlimitations}

Although \textsc{GGchem}'s enormous chemical network accounts for a large number of mineral species, it does not account for mineral solid solution, which for certain phases is strongly favoured in their thermodynamic stability over pure-phase endmember minerals.
For example, olivine (any member of the solid solution (Mg,Fe)$_2$SiO$_4$) is represented within \textsc{GGchem} only as either pure forsterite (Mg$_2$SiO$_4$) or pure fayalite (Fe$_2$SiO$_4$).
As such, the mineralogies that we have inferred here will not be precisely those that would be stable under the respective equilibrium conditions.
This is however likely to be a second-order effect, with our results at least providing an accurate approximation to the true equilibrium mineralogies.

This also highlights consequences for the range of conditions over which surface--atmosphere equilibrium will dominate over other processes, in particular the melting of the crust.
At high temperatures, \textsc{GGchem} accounts for the melting of pure solid minerals into their liquid form, at temperatures in excess of $1700\,\text{K}$.
However, eutectic melting means that melts will form at lower temperatures than is the case for pure phases.  This enables planets to transition to the the lava world scenario of Figure~\ref{fig:regimes} at lower temperatures than predicted here, allowing mass transfer and chemical reaction between reservoirs to occur much more rapidly.

As such, although we have limited the scope of this work to surface temperatures below about $1700\,\text{K}$, the assumption of solid--gas equilibrium being the dominant force in setting the atmospheric and surface compositions becomes less valid at the upper end of this temperature range.

\section{Conclusions} \label{conclusion}

We have investigated the thermochemical equilibrium between the surface and the atmosphere of warm rocky exoplanets, finding a boundary in surface pressure--temperature space which delineates the stability regimes of the major gases in the atmosphere, concurrent with the stability of several minerals.
This potentially enables observational inferences of the presence or absence of specific minerals on the surfaces of warm rocky exoplanets and of their surface pressure temperature conditions, using contemporary techniques such as transmission spectroscopy for analysing exoplanetary atmospheres.
The inferences we have outlined could provide a complementary method to identifying exoplanet mineralogy, alongside emission spectroscopy.

The surface--atmosphere regime boundary we identify is not entirely sharp, but rather a narrow intermediate region of pressure--temperature space where the mineralogical and atmospheric chemistries transition from one regime to the other.
Although this somewhat blurs the picture for a small range of surface conditions, we argue that the correlation that we have found between the two atmospheric/mineralogical regimes is sufficiently stark to enable unprecedented inferences of the surface mineralogy of warm terrestrial exoplanets.

\section*{Acknowledgements}

The authors thank the Institute of Astronomy, University of Cambridge, for the funding of the internship within which this work was conducted.
S.J. thanks the Science and Technology Facilities Council (STFC) for the PhD studentship (grant reference ST/V50659X/1).
We thank the anonymous reviewer for their constructive report.
We also thank Aprajit Mahajan for his suggestions for viable compositional ranges.

\section*{Data Availability}

The \textsc{GGchem} input and abundance files, the resulting output files, and data analysis code will be made available upon acceptance of the manuscript at \url{https://github.com/xbyrne/sai}.



\bibliographystyle{mnras}
\bibliography{main_bib}




\appendix

\section{Depth of Rock in Equilibrium with Atmosphere}

\label{app:layerdepth}
Whilst on rocky planets the rock--reservoir, by definition, vastly exceeds the atmospheric reservoir in mass, the majority of this is buried and not able to participate in atmospheric chemistry.
In the context of surface--atmosphere interaction on Earth, this question is central to the operation of the carbon cycle, and whether climate regulation is rock-limited or not \citep[e.g.,][]{graham2020thermodynamic}.
For the Venus-like case investigated here, the key question is whether there is sufficient unsulfidated calcium in the upper crust to absorb all of the atmospheric \ce{SO2} present in the atmosphere at high temperatures; this paper shows this to be the primary reaction at play.

One could envisage a scenario where there would only be enough anorthite and diopside to desulfidate the atmosphere with a crust `skin depth' of many kilometres.
In such a scenario, it would be unlikely that the atmospheric regimes we have identified would be able to occur in practice; rock kilometers deep in the crust would likely be shielded from atmospheric interaction on the time-scales of hundreds of millions of years. 
However, on sub--metre scales, it is reasonable that fractures and permeability of the crust would allow it to be in effective contact with the atmosphere to this depth, facilitating the reactions needed to reach equilibrium.

The number of sulfur atoms in a given atmospheric column is given by $N_{\ce{S}}$, where

\begin{equation}
p_0
= \int \rho g \dd{z}
= \mu g \int n \dd{z}
= \frac{\mu g}{x_{\ce{S}}} \int n_{\ce{S}} \dd{z}
= \frac{\mu g N_{\ce{S}}}{x_{\ce{S}}}
\end{equation}
where $x_{\ce{S}}$ is the combined mixing ratio of all sulfur-bearing gases.
As the \ce{Ca}:\ce{S} stoichiometry of reactions~\ref{eq:diopside} and \ref{eq:anorthite} is 1:1, each atmospheric sulfur atom could in principle combine with one calcium atom in the formation of \ce{CaSO4}.
The skin depth is therefore:
\[
L_{\ce{Ca}}
= \frac{N_{\ce{S}}}{n_{\ce{Ca}}}
= \frac{x_{\ce{S}} p_0}{\mu g n_{\ce{Ca}}}
\]
where $n_{\ce{Ca}}$ is the number density of calcium atoms in the solids.
We find that over the pressure--temperature conditions investigated, $L_{\ce{Ca}}<15\,\text{cm}$.
Over this range we would expect fractures in the crust to enable equilibrium to be reached with the lower atmosphere.

\section{Elemental Composition of Venus}
\label{app:elements}

The elemental composition of Venus that we have used throughout is estimated using section 3.2 of \cite{rimmer21}.
This procedure uses oxide mass fractions as measured by the \textit{Vega 2} lander, and assumes an SO$_2$ VMR of $150\,\text{ppm}$ at the surface.

\begin{table}
\caption{
Estimated elemental composition of Venus' crust.
This table uses the common astronomical convention 
$\epsilon_\text{X} = 12 + \log_{10}(n_\text{X} / n_\text{H})$, where $n_\text{X}$ is the number density of nuclei of element X.
The procedure for determining these abundances is described in section 3.2 of \citet{rimmer21}
which estimates Venus' surface composition using \textit{Vega 2} measurements of oxide mass fractions \citep{surkov86}.
}
\centering
\begin{tabular}{|c|c|c|c|}
\hline
Element & $\epsilon$ & Element & $\epsilon$\\
\hline
H  & 12.0000 & Mg & 18.5693 \\
C  & 16.0985 & Ca & 18.2402 \\
N  & 14.9591 & Al & 18.6107 \\
O  & 19.5591 & Na & 17.9238 \\
F  & 10.3444 & K  & 16.4410 \\
S  & 17.8827 & Ti & 16.5126 \\
Cl & 9.73721 & He & 11.1931 \\
Fe & 18.1441 & Ne & 10.8129 \\
Mn & 16.4092 & Ar & 11.9591 \\
Si & 18.9942 &&\\
\hline
\end{tabular}
\label{tab:elements}
\end{table}


\bsp	
\label{lastpage}
\end{document}